# Multi-objective topology optimization of heat transfer surface using level-set method and adaptive mesh refinement in OpenFOAM

Di Chen[1], Prashant Kumar[1], Yukinori Kametani[2], Yosuke Hasegawa[1]*

*Corresponding author: ysk@iis.u-tokyo.ac.jp
1 Institute of Industrial Science, The University of Tokyo
2 School of Science and Technology, Meiji University

**Abstract**

The present study proposes a new efficient and robust algorithm for multi-objectives topology optimization of heat transfer surfaces to achieve heat transfer enhancement with a less pressure drop penalty based on a continuous adjoint approach. It is achieved with a customized OpenFOAM solver, which is based on a volume penalization method for solving a steady and laminar flow around iso-thermal solid objects with arbitrary geometries. The fluid-solid interface is captured by a level-set function combined with a newly proposed robust reinitialization scheme ensuring that the interface diffusion is always kept within a single local grid spacing. Adaptive mesh refinement is applied in near-wall regions automatically detected by the level-set function to keep high resolution locally, thereby reduces the overall computational cost for the forward and adjoint analyses. The developed solver is first validated in a drag reduction problem of a flow around a two-dimensional cylinder at the Reynolds numbers of 10 and 40 by comparing reference data. Then, the proposed scheme is extended to heat transfer problems in a two-dimensional flow at the Prandtl number of 0.7 and 6.9. Finally, three-dimensional topology optimization for multi-objective problems is considered for cost functionals with different weights for the total drag and heat transfer. Among various solutions obtained on the Pareto front, 4.0% of heat transfer enhancement with 12.6% drag reduction is achieved at the Reynolds number of 10 and the Prandtl number of 6.9. Moreover, the optimization of a staggered pin-fin array demonstrates that the optimal shapes and arrangement of the fins strongly depend on the number of rows from the inlet. Specifically, the pin-fins in the first and third rows extended in the upstream direction further enhance heat transfer, while the fins in the second row vanish to reduce pressure loss.

## 1. Introduction

Shape/topology optimization of fluid-solid interfaces for enhancing heat and mass transfer with a less pressure drop penalty is a key technology in the efficient use of energy in the civilian and industrial sectors. With the rapid advancements in additive manufacturing technology and novel functional materials in the past decades, topology optimization (TO) has attracted much attention as a tool to obtain new design concepts of heat transfer surfaces [1]. Significant improvements in thermal and hydraulic performances from conventional heat transfer surfaces have been reported through shape/topology optimization [2].

Existing studies of heat sinks TO have considered either hydraulic resistance or thermal dissipation [5]-[10], or multiple objectives taking into account both of them [2]-[4].

Shape/topology optimization algorithms are generally classified into gradient-based and gradient-free methods. The gradient-based methods calculate the gradient (or sensitivity) of a prescribed objective function with respect to design variables describing the geometry of a solid object. The design variables are iteratively updated by the sensitivity information. The update of the design variables is conducted with different algorithms such as the steepest descent method [5], [11], [12], Method of Moving Asymptotes (MMA) [3], [4], [13], Sequential Linear/Quadratic Programming (SLP/SQP) [2], [10], etc., as reviewed by Dbouk [1]. On the other hand, gradient-free methods such as genetic algorithms [14], [15], particle swarm optimization [16], and Bayesian optimization [17] have the potential to find the global minimum, since it does not rely on the local gradient. However, they often require a large number of trials, so their applications are limited to optimization problems for design variables with relatively small degrees of freedom around the order of ten.

Gradient-based algorithms have been developed for shape optimization since the pioneering study by Pironneau [18] for obtaining the optimal airfoil profile with the minimum drag based on an adjoint method. The unique advantage of the adjoint method is that the sensitivity of each surface element to a prescribed cost function can be obtained by solving the adjoint equation of the forward problem only once. In shape optimization, a fluid-solid boundary is clearly defined and only the fluid region is solved with boundary-fitted mesh. However, the requirement of the remeshing for updated fluid-solid boundaries [7] significantly increases the overall computational cost. Therefore, most existing shape optimization studies consider two-dimensional problems with relatively coarse mesh, and significant efforts have been devoted to the development of numerical schemes and optimization algorithms [1]. For instance, Kim et al. [5] proposed a method for reducing design variables to be optimized in the adjoint analysis. More recently, Chen et al. [19] employed deep neural networks for solving the flow fields, while a PyTorch package for calculating the sensitivity information.

In topology optimization, the solid geometry is usually represented by the volume ratio of the solid in each computational cell. This allows to handle topology changes such as connection of isolated solids and generation of a hole in solid in a straightforward manner. Meanwhile, since the fluid-solid interface is not explicitly defined as in the shape optimization, but has to be reconstructed from the volume information, it poses significant challenges in resolving momentum and thermal boundary layers formed near the wall. The solid region can be parameterized by a few design variables [20] or the porosity or material distribution defined at each grid point [5], [11]. The latter approach dramatically increases the degrees of freedom of design variables, while it enables more flexible topology changes. Solid Isotropic Material with Penalization (SIMP) method is one of such homogenization approaches first introduced by Bendsøe and Kikuchi [25] for structural optimal design, in which a smooth, but rapid transition between void and solid is introduced by penalizing intermediate densities. Numerical filtering is also often employed to control the convexity of the penalization coefficients [8], while their tuning is non-trivial. For flow problems, Borrvall and Petersson [26] solved the TO problem in Stokes flow by introducing a density function to distinguish the fluid and solid phases. The density approach has inspired a lot of TO studies for the optimal designs of heat sinks [2], [4], [10], [13]. However, the conventional density approaches often result in complex porous structures, which might be optimal, but quite difficult to fabricate.

To overcome such difficulties, a level-set method (LSM) has a great advantage for capturing the fluid-solid interface with topology change. It was first introduced by Osher and Sethian [27] and has widely been used for describing the evolution of the interface between two phases. In this approach, the fluid-solid interface is implicitly represented by a zero contour of the level-set function, which is defined at each grid point as the signed distance from the nearest fluid-solid interface. Leveraging its unique properties, LSM-based TO possesses three main advantages over the density approach as follows: First, the clear fluid-solid interface described by LSM significantly improves the accuracy of the forward and adjoint simulations. This is particularly critical in flow simulations since it is important to resolve the velocity and thermal boundary layers formed near the fluid-solid interface. Second, LSM also improves the convergence behavior and computational speed in TO, as the sensitivity is only computed in a transition region of the interface rather than the entire domain in density approaches. Third, the transition region in LSM is uniformly distributed around the fluid-solid interface based on the signed distance, and its thickness can be fixed by a user. Accordingly, the phase indicator and all the physical properties are smoothly changed between those of fluid and solid in the transition region.

In LSM-based TO, once the interface is updated based on the sensitivity obtained from the adjoint analysis, it is necessary to reinitialize the level-set function to maintain its feature as the signed distance function (SDF). This procedure is generally called reinitialization, which is often realized by solving an additional partial differential equation as originally introduced by Sussman et al [28]. Ideally, during the reinitialization procedure, the zero-contour of the level-set function corresponding to the interfacial location should not be modified, while only the surrounding level-set function is corrected. Due to numerical diffusion introduced in conventional reinitialization methods [28], [29], however, maintaining the interfacial location during the reinitialization remains a difficult task [9], [30]. Therefore, with the existing LSM-based TO, smaller-scale updates of the geometry tend to be more easily smeared during the reinitialization, and this significantly limits the applications of LSM in topology optimization.

Some recent studies propose novel schemes to skip the reinitialization procedure by introducing an additional regularization term [31] or restricting the range of the level-set function close to the interface [9]. However, both schemes still suffer from numerical diffusion of the fluid-solid interface, and thereby the ambiguity in the interfacial location remains [8], [23], [32]. More recently, Zhang et al. [33] introduced an internal iteration procedure to correct the interfacial location to cancel the numerical diffusion during the reinitialization. However, such a technique has not been combined with TO algorithm yet. Furthermore, introducing the internal loop in the reinitialization procedure could be cumbersome. In the present study, we propose a novel strategy to maintain the interfacial location within a local one-grid spacing during the reinitialization procedure without introducing the iterative process proposed by Zhang et al. [33]. The proposed algorithm is implemented in the open-source CFD solver, OpenFOAM, together with an original adjoint code for TO. The Adaptive Mesh Refinement (AMR) function originally integrated with the OpenFOAM repository allows to automatically generate multi-step mesh refinement based on the level-set function, i.e., the distance from the interface. As a result, this significantly improves the representation of the fluid-solid interface and enables more accurate computation of fluid flow and associated heat transfer between fluid and solid.

The purposes of the present study are as follows: We propose a new reinitialization procedure for LSM and combine it with a customized OpenFOAM TO solver. Then, the developed code is verified in a drag reduction problem of a single cylinder placed in a uniform through comparison with reference data obtained by sharp-interface approaches [5], [19]. Second, we extend the current optimization algorithm to multi-objective optimization problems aiming at simultaneous achievement of heat transfer enhancement and drag reduction. Finally, we apply the current algorithm to a three-dimensional single pin fin and also an array of pin fins to show the advantages of LSM for handing complex three-dimensional shapes and topology changes.

## 2. Methodology

### 2.1. Level-set function for representing fluid-solid interface

In the present study, we represent an arbitrary complex fluid-solid interface with a level-set function $\phi_0$ [28], [32]. The level-set function is a signed distance function (SDF) from the nearest fluid-solid interface and defined to be positive and negative inside the solid and fluid regions, respectively. Therefore, the 3D fluid-solid interface can be expressed by the iso-surface of $\phi_0 = 0$ as shown in Fig. 1(a). Once the level-set function is given, the phase indicator $\phi$ [8], [23] is defined by the following formula:

$$\phi = \begin{cases} 0 & \phi_0 < -\delta_t \\ \frac{1}{2}\left[\sin\left(\frac{\pi\phi_0}{2\delta_t}\right) + 1\right] & -\delta_t \leq \phi_0 \leq \delta_t \\ 1 & \phi_0 > \delta_t \end{cases} \quad (1)$$

Hence, $\phi$ smoothly changes from zero (for fluid) to unity (for solid) in the form of the sine function across the fluid-solid transition region $-\delta_t \leq \phi_0 \leq \delta_t$. Such a diffused-interface approach allows to embed a complex fluid-solid interface in a Cartesian grid system. Here, $\delta_t$ is the one-sided transition layer thickness defined by a user. The schematic of the spatial distributions of the level-set function and the phase indicator is shown in Fig. 1(b). According to Eq. (1), the derivative of the phase indicator with respect to the level-set function, $d\phi/d\phi_0$, has a non-zero value only within the transition layer as shown in Fig. 1(b), and it approaches to the delta function in the 3D space with decreasing the transition layer thickness $\delta_t$. Ideally, $\delta_t$ should be sufficiently thin. In the present study, it is determined so that the transition region is distributed over a few local grid points near the interface, which can effectively be reduced by adaptive mesh refinement as will be introduced later.

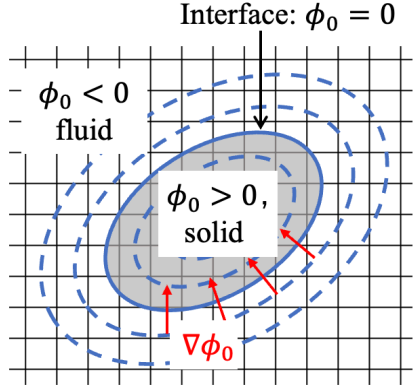
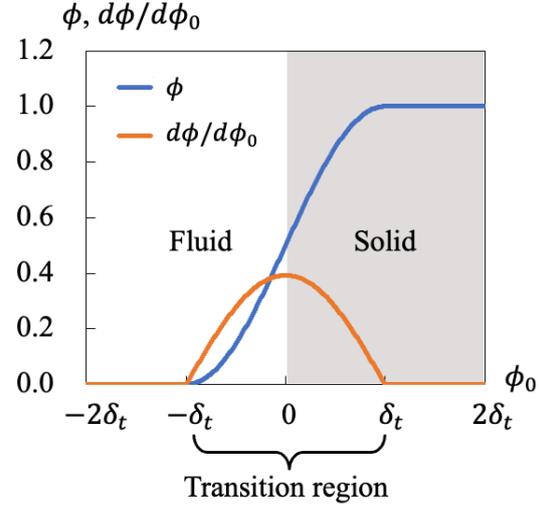

**Figure 1.** (a) Iso-surfaces of $\phi_0$ in the Cartesian grids showing solid region ($\phi_0 > 0$), fluid region ($\phi_0 < 0$), and interface ($\phi_0 = 0$), red arrows indicate $\nabla \phi_0$ corresponding to the interface-normal vector with a unit length. (b) Schematic of the distributions of the phase indicator $\phi$ and its derivative $d\phi/d\phi_0$ as function of $\phi_0$ near the interface ($\phi_0 = 0$).

When an initial shape is embedded in a Cartesian mesh, the level-set function $\phi_0$ is initially set to be -1 and 1 in the fluid and solid regions, respectively. In order for the level-set function to be a signed distance function, the Eikonal equation $|\nabla \phi_0| = 1$ has to be satisfied throughout the computational domain, while the value of the level-set function is zero on the fluid-solid interface. To achieve it, the reinitialization of the level-set function is commonly conducted by solving the following pseudo-time-dependent partial differential equation [28], [29]:

$$\frac{\partial \phi_0}{\partial \tau} + \text{sign}(\phi_0)(1 - |\nabla \phi_0|) = 0, \quad (2)$$

where

$$\text{sign}(\phi_0) = \begin{cases} -1 & \text{if } \phi_0 < 0 \\ 0 & \text{if } \phi_0 = 0 \\ 1 & \text{if } \phi_0 > 0 \end{cases}, \quad (3)$$

and $\tau$ is a pseudo time. From Eq. (2), it can be seen that $|\nabla \phi_0| = 1$ is satisfied when $\phi_0$ is converged. In solving Eq. (2), it is commonly converted into the form of an advection-diffusion equation with unit velocity $\widetilde{\boldsymbol{u}} = \text{sign}(\phi_0) \nabla \phi_0 / |\nabla \phi_0|$ as follows:

$$\frac{\partial \phi_0}{\partial \tau} + (\widetilde{\boldsymbol{u}} \cdot \nabla) \phi_0 = \text{sign}(\phi_0) + D \nabla^2 \phi_0, \quad (4)$$

where $D$ is an artificial diffusion coefficient for suppressing numerical instability. Even though the numerical diffusion is kept small during a single reinitialization procedure, it leads to significant smoothing of the fluid-solid interface after multiple iterations, especially near a sharp corner [9], [23], [30]. Figures (2a) and (2b) show how the conventional reinitialization scheme smoothens the original square shape when the iteration number increases from $N_r = 5$ to 50. The pseudo time for reinitialization is $\tau = N_r \Delta\tau$, where $\Delta\tau$ (pseudo time step) is determined by the maximum Courant number of 0.3. Here, the white broken line represents the original square, whereas the yellow line corresponds to the iso-surface of $\phi_0 = 0$ after the reinitialization.

To overcome this problem, we propose a novel simple algorithm for the reinitialization of $\phi_0$. We assume that a level-set function before applying reinitialization has the following properties: First, it is zero along the fluid-solid interface. It also has positive and negative values in solid and fluid regions, while it may not be the signed distance function. This situation is quite common. For example, most software including OpenFOAM used in the present study have a function to read geometry data such as a Standard Triangulated Language (STL) file and assign 1 and -1 at each grid point depending on whether the location is inside or outside a solid region. Also, the above properties are maintained for a level-set function updated based on the adjoint analysis explained in Sec. 2.3.

In the present reinitialization procedure, we first estimate the distances from the fluid-solid interface at adjacent grid points around the interface, $\widetilde{\phi_0}$, from a level-set function before applying the reinitialization as follows:

$$\widetilde{\phi_0} = \frac{\phi_0}{|\nabla \phi_0|}. \tag{5}$$

If the initial $\phi_0$ is zero at the fluid-solid interface and also smoothly changes from negative to positive values across the fluid-solid interface, Eq. (5) provides the signed distance from the interface in the second-order accuracy. Then, we solve the standard reinitialization partial differential equation (4), while the value of $\widetilde{\phi_0}$ is always assigned to each adjacent grid point during the reinitialization. Since the values of $\phi_0$ in proximity to the interface are maintained, the interfacial location always stays within a single grid spacing during the reinitialization. It can be seen that the present algorithm significantly suppresses the smoothing the interface even after 50 iterations, especially near the corners (compare Figs. (2b) and (2c)).

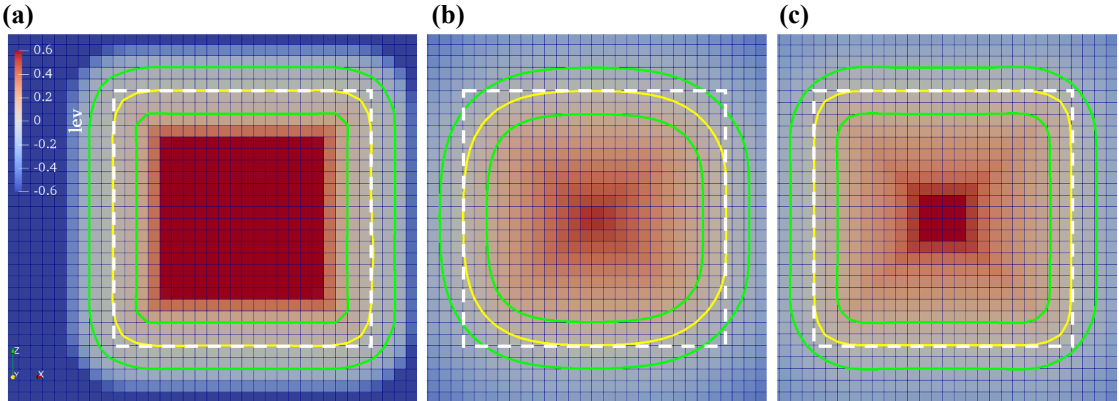

**Figure 2.** Level-set functions calculated by the standard reinitialization algorithm after iterations of (a) $N_r = 5$ and (b) $N_r = 50$, and (c) by the newly proposed reinitialization algorithm iterations of $N_r = 50$. Yellow and green isolines denote $\phi_0 = 0$, and $\pm 0.1$, respectively. White dashed lines are the original square shape.

We note that the present reinitialization scheme has several advantages over the two-iteration scheme proposed by Zhang et al. [33]. In the present approach, the distance from the interface on an adjacent grid point is calculated only once from the initial level-set function before reinitialization, and this value is assigned to the adjacent grid point after each time step of the reinitialization. Since the algorithm proposed by Zhang et al. [33] requires an iterative procedure at each reinitialization time step, our approach is simpler and cheaper, while it is ensured that the location of the interface can be kept within a local single grid spacing during the reinitialization.

### 2.2. Governing equations

In the present study, we consider an incompressible fluid and treat the temperature as a passive scalar. We consider a low Reynolds number flow, so that the velocity and thermal fields are assumed to be steady and laminar. Accordingly, the governing equations for momentum, mass, and energy conservations are expressed by the following non-dimensionalized forms:

$$\frac{\partial(u_j u_i)}{\partial x_j} + \frac{\partial p}{\partial x_i} - \frac{1}{Re}\frac{\partial^2 u_i}{\partial x_j \partial x_j} + \eta_u \phi (u_i - u_i^s) - G_i = 0, \tag{6}$$

$$\frac{\partial u_i}{\partial x_i} = 0, \tag{7}$$

$$\frac{\partial(u_j \theta)}{\partial x_j} - \frac{1}{PrRe}\frac{\partial^2 \theta}{\partial x_j \partial x_j} + \eta_\theta \phi (\theta - \theta^s) - S = 0, \tag{8}$$

where $x_i$ and $u_i$ are the coordinate and the corresponding velocity component in the $i$-th direction, whilst $p$ and $\theta$ are the static pressure and the temperature, respectively. Here, $G_i$ represents the body force acting on the fluid such as gravity and electromagnetic forces, while $S$ represents a heat source. The Reynolds and Prandtl numbers are defined as $Re = \rho U_\infty L_{ref}/\mu$, and $Pr = \mu C_p/\lambda_f$, respectively, where $\rho$, $\mu$, $C_p$ and $\lambda_f$ represent the density, dynamic viscosity, specific heat capacity, and thermal conductivity of the fluid, respectively. Equations (6-8) are nondimensionalized by the characteristic length-scale $L_{ref}$ and the velocity-scales $U_\infty$.

We note that the present scheme embeds an arbitrary solid structure in the Cartesian grid system, and Eqs. (6-8) are solved at all the grid points in both fluid and solid regions. The second last term on the right-hand side of Eqs. (6) and (8) are artificial body force and heat source, respectively. They are introduced to impose the no-slip and iso-thermal boundary conditions at a fluid-solid interface using the volume penalization method (VPM) [34]. These terms are null in the fluid region and only act to make the velocity and thermal fields to converge to the target values, $u_i^s$ and $\theta^s$, inside the solid. In the present study, $u_i^s$ and $\theta^s$ are set to be zero, so that no-slip and iso-thermal conditions are satisfied on the fluid-solid interface. The penalization coefficients $\eta_u$ and $\eta_\theta$ are set sufficiently large, i.e., $\eta_u = \eta_\theta = 10^6$, so that the velocity and thermal boundary conditions are satisfied while maintaining numerical stability.

## 2.3. Adjoint analysis

2.3.1. Cost function

In the present adjoint-based TO algorithm, a fluid-solid interface, which is represented by a zero-contour of the level-set function, is updated so as to minimize a prescribed cost functional. Since we aim to enhance heat transfer with a less pressure loss, we consider a multi-objective cost functional $J$ expressed by the linear sum of the drag force ($F_i$), the heat transfer on ($Q$), the bulk mean velocity ($U_b$), the bulk mean temperature ($\Theta_b$), the solid region volume ($V$), and the interface area ($A$) as follows:

$$J = C_{F_i}F_i + C_Q Q + C_U U_b + C_\Theta \Theta_b + C_V V + C_A A, \tag{9}$$

where $C_{F_i}, C_Q, C_U, C_\Theta, C_A$, and $C_V$ correspond to the individual weight coefficients.

The drag force (in the $i$-th direction) acting on the solid object in the fluid domain can be calculated by the volume integral of the artificial body force in Eq. (6) as

$$F_i = \int_\Omega \phi \eta_u (u_i - u_i^s) dV. \tag{10}$$

Similarly, the total heat transfer at the fluid-solid interface can also be calculated as

$$Q = \int_\Omega \phi \eta_\theta (\theta - \theta^s) dV. \tag{11}$$

The bulk mean velocity $U_b$ and the bulk mean temperature $\Theta_b$ are respectively defined as follows:

$$U_b = \langle u_1 \rangle_V \equiv \frac{1}{V_\Omega} \int_\Omega u_1 dV, \Theta_b = \langle \theta \rangle_V \equiv \frac{1}{V_\Omega} \int_\Omega \theta dV, \tag{12}$$

where $u_1$ is the streamwise velocity component, and $V_\Omega$ is the total volume of the computational domain $\Omega$. The bracket $\langle \ \rangle_V$ indicates the volume integral within the computational domain $\Omega$.

The present problem can be considered as the minimization of the cost functional (9) under the constraints of Eqs. (6-8). This is equivalent to minimizing the following Hamiltonian:

$$H = J - \langle u_i^* \cdot Eq.(6) - p^* \cdot Eq.(7) + \theta^* \cdot Eq.(8) \rangle_V$$
$$+ G_i^* \cdot (\langle u_i \rangle_V - U_i^{tar}) + S^* \cdot (\langle \theta \rangle_V - \Theta_i^{tar}) + \phi^* \cdot (\langle \phi \rangle_V - \Phi^{tar}), \tag{13}$$

where $u_i^*, p^*$, and $\theta^*$ are Lagrange multipliers for the individual constraints. They are also referred to as the adjoint variables. $U_i^{tar}$, $\Theta_i^{tar}$, and $\Phi^{tar}$ respectively represent constraints keeping the bulk mean velocity, temperature, and the solid volume constant. When such a constraint is not present, the corresponding term in the last three terms of Eq. (13) can be removed by setting $G_i^*$, $S^*$ or $\phi^*$ to be zero.

2.3.2. Adjoint equations and sensitivity

In the gradient methods [1]-[5], [10]-[13], a design variable is updated based on its sensitivity to a cost functional. In the present study, the design variable is the phase indicator $\phi$, which indicates the local solid volume ratio inside the computational domain. Assuming an arbitrary infinitesimal change ($\phi'$) in the design variable, and then applying the Fréchet differential to $H$ results in

$$H' = \frac{DH}{D\phi}\phi' = \langle\phi'\{C_{F_i}V_\Omega\eta_u(u_i - u_i^s) + C_Q V_\Omega\eta_\theta(\theta - \theta^s) - \eta_u(u_i - u_i^s)u_i^* - \eta_\theta(\theta - \theta^s)\theta^*$$

$$+ C_V - C_A \nabla^2\phi/|\nabla\phi| + \phi^*\}\rangle_V \tag{14a}$$

$$+ \left\langle u_i'\left\{u_j\left(\frac{\partial u_j^*}{\partial x_i} + \frac{\partial u_i^*}{\partial x_j}\right) - \frac{\partial p^*}{\partial x_i} + \frac{\partial}{\partial x_j}\left(\frac{1}{Re}\frac{\partial u_i^*}{\partial x_j}\right) + \eta_u\phi(C_{F_i}V_\Omega - u_i^*)\right.\right.$$

$$\left.\left. + G_i^* + \theta\frac{\partial\theta^*}{\partial x_i} + C_U\delta_{i1}\right\}\right\rangle_V \tag{14b}$$

$$+ \left\langle p'\frac{\partial u_i^*}{\partial x_i}\right\rangle_V \tag{14c}$$

$$+ \left\langle \theta'\left\{u_j\frac{\partial\theta^*}{\partial x_j} + \frac{\partial}{\partial x_j}\left(\frac{1}{Pe}\frac{\partial\theta^*}{\partial x_j}\right) + \eta_\theta\phi(C_Q V_\Omega - \theta^*) + S^* + C_\Theta\right\}\right\rangle_V \tag{14d}$$

$$+ G_i'\langle u_i^*\rangle_V + S'\langle\theta^*\rangle_V \tag{14e}$$

$$- \left\langle\frac{\partial}{\partial x_j}\left\{u_i^*(u_j u_i)' + p'u_j^* - p^* u_j' - \frac{1}{Re}\left(u_i^*\frac{\partial u_i'}{\partial x_j} - u_i'\frac{\partial u_i^*}{\partial x_j}\right)\right.\right.$$

$$\left.\left. + \theta^*(u_j\theta)' - \frac{1}{PrRe}\left(\theta^*\frac{\partial\theta'}{\partial x_j} - \theta'\frac{\partial\theta^*}{\partial x_j}\right)\right\}\right\rangle_V \tag{14f}$$

Here, the terms (14b), (14c), and (14d) vanish when the adjoint velocity and temperature satisfy the following adjoint equations:

$$-u_j\left(\frac{\partial u_j^*}{\partial x_i} + \frac{\partial u_i^*}{\partial x_j}\right) = -\frac{\partial p^*}{\partial x_i} + \frac{\partial}{\partial x_j}\left(\frac{1}{Re}\frac{\partial u_i^*}{\partial x_j}\right) + \eta_u\phi(C_{F_i}V_\Omega - u_i^*) + G_i^* + \theta\frac{\partial\theta^*}{\partial x_i} + C_U\delta_{i1}, \tag{15}$$

$$\frac{\partial u_i^*}{\partial x_i} = 0, \tag{16}$$

$$-u_j\frac{\partial\theta^*}{\partial x_j} = \frac{\partial}{\partial x_j}\left(\frac{1}{PrRe}\frac{\partial\theta^*}{\partial x_j}\right) + \eta_\theta\phi(C_Q V_\Omega - \theta^*) + S^* + C_\Theta. \tag{17}$$

The term (14e) also vanishes when $G_i$ and $S$ are kept constant. The last term (14f) can be transformed to the surface integral on the domain boundary as follows:

$$-\frac{1}{V_\Omega}\int_{d\Omega}\left\{n_j(u_i^* u_i) + u_j^*(u_i n_i) - n_j p^* + \frac{1}{Re}\left(n_i\frac{\partial u_j^*}{\partial x_i}\right) + \theta^*\theta n_j\right\}u_j' dS \tag{18a}$$

$$+ \frac{1}{V_\Omega}\int_{d\Omega}\frac{1}{Re}n_j\left\{\left(\frac{\partial u_i'}{\partial x_j}\right)u_i^*\right\} dS \tag{18b}$$

$$- \frac{1}{V_\Omega}\int_{d\Omega}(n_j u_j^*)p' dS \tag{18c}$$

$$- \frac{1}{V_\Omega}\int_{d\Omega}\left\{\theta^*(u_j n_j) + \frac{1}{PrRe}\left(n_j\frac{\partial\theta^*}{\partial x_j}\right)\right\}\theta' dS \tag{18d}$$

$$+ \frac{1}{V_\Omega}\int_{d\Omega}\frac{1}{PrRe}n_j\left\{\left(\frac{\partial\theta'}{\partial x_j}\right)\theta^*\right\} dS. \tag{18e}$$

The boundary conditions for solving the adjoint equations (15-17) are determined by cancelling all these boundary terms from (18a) to (18e). As the adjoint boundary conditions also depend on the physical boundary conditions for solving Eq. (6-8), they will be provided for specific cases in Sec. 3.2 and 3.3.

Eventually, only the first term (14a) remains as the sensitivity to the change of $\phi$ on the right-hand side of Eq. (14). This reduces to

$$H' = \langle \phi' \{ (C_{F_i} V_\Omega - u_i^*) \eta_u (u_i - u_i^s) + (C_Q V_\Omega - \theta^*) \eta_\theta (\theta - \theta^s) + C_V - C_A \nabla^2 \phi / |\nabla \phi| + \phi^* \} \rangle_V \qquad (19)$$

Here, $\phi'$ can be related to the change in the level-set function, i.e., $\phi_0'$ by the chain rule as

$$\phi' = \phi_0' \frac{d\phi}{d\phi_0}, \qquad (20)$$

where $d\phi/d\phi_0$ has a non-zero value only within the transition region from 0 to 1 as shown in Fig. 1(b). By applying the steepest descent method for updating $\phi_0$, it is guaranteed that the Hamiltonian can always be reduced if $\phi_0$ is updated by the following formula:

$$\phi_0' = -\alpha \frac{d\phi}{d\phi_0} \left\{ (C_{F_i} V_\Omega - u_i^*) \eta_u (u_i - u_i^s) + (C_Q V_\Omega - \theta^*) \eta_\theta (\theta - \theta^s) + C_V - \frac{C_A \nabla^2 \phi}{|\nabla \phi|} + \phi^* \right\}, \qquad (21)$$

where $\alpha$ represents the rate of the update and is set to be unity in the present study.

## 2.4. Numerical implementation and overall procedures

In this study, the overall TO process including forward and adjoint simulations is conducted in the open-source CFD toolbox, OpenFOAM (Version 8). An arbitrary solid geometry is embedded in a Cartesian grid system by using the level-set function. Adaptive Mesh Refinement (AMR) is then applied to refine the mesh in multiple levels as approaching to the fluid-solid interface. It is done automatically based on the level-set function corresponding to the signed distance from the nearest interface. In each refinement, a numerical cell close to the interface is divided in half, so the finest grid spacing near the interface is $1/2^n$ of the original grid size, where $n$ indicates the maximal refinement level. In the present study, $n = 2$ is used, since this refinement level is found to be sufficient to obtain converged solutions as shown in Sec. 3.1. In solving the forward governing equations (6) to (8) and also the adjoint equations (15) to (17), the Gauss linear scheme is applied for gradient and Laplacian calculation, while the Gauss upwind scheme is applied for calculation of the divergence.

Figure 3 shows the flow chart of the overall TO process adopted in the present study. It consists of the following procedures:

Step 1: Initialize level-set function by reading STL file of an initial geometry, and solve the reinitialization equation (4) with our new algorithm explained in Sec. 2.1.

Step 2: Solve the forward governing equations (6-8) for steady velocity and thermal fields.

Step 3: Solve the steady adjoint velocity and temperature equations (15-17).

Step 4: Calculate the sensitivity of the design variables to the cost functional by Eq. (21) based on the forward and adjoint velocity and thermal fields obtained in Steps 2 and 3.

Step 5: Update the level-set function by $\phi_0 = \phi_{0,old} + \Delta\phi_0$, and also the phase indicator $\phi$ based on the updated $\phi_0$.

Step 6: Reinitialize the level-set function by the new algorithms described in Sec. 2.1.

Step 7: Apply adaptive mesh refinement based on the reinitialized level-set function.

Step 8: Evaluate the cost function (9). If the convergence criterion defined below is not met, return to Step 2 using the updated design variable $\phi_0$ (setting $\phi_{0,old} = \phi_0$).

The convergence criterion for the present TO algorithm is defined as

$$\left| J_n - \sum_{n-\delta_{it} \leq k < n} J_k / \delta_{it} \right| < \epsilon, \tag{22}$$

where $n$ is the iteration number. In this study, the iteration interval $\delta_{it}$ for checking the convergence state and the criterion value $\epsilon$ are set to be 50 and $10^{-4}$, respectively.

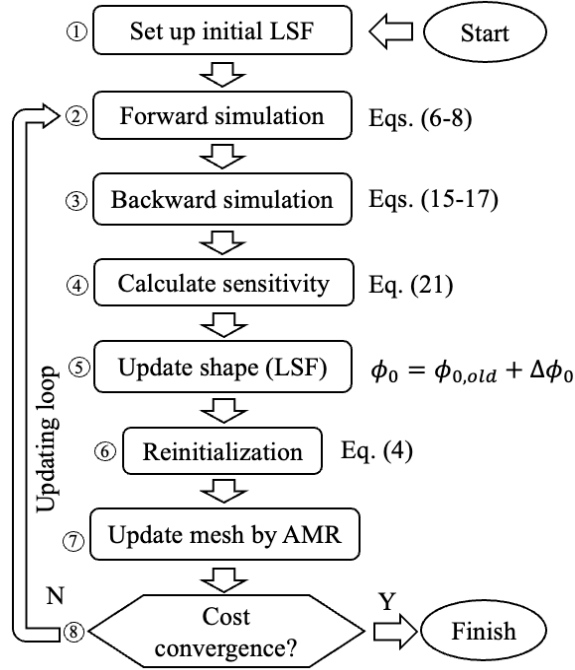

**Figure 3.** Flowchart of TO process using LSM in OpenFOAM

3. **Results and discussion**

This section presents the optimal designs for multi-objective topology optimization in drag reduction and heat transfer enhancement problems. The section is organized as follows. Firstly, 2D optimization is performed to solely reduce the drag on a circular cylinder at *Re* values of 10 and 40 in Sec. 3.1. These benchmark problems are used to verify the current optimization algorithm. In addition, a grid convergence study in the forward simulation is reported to verify the present grid resolution. In Sec. 3.2, the present algorithm is extended to an optimization problem for heat transfer enhancement at Prandtl numbers of 0.7 and 6.9, which correspond to air and water under the standard condition, respectively. Finally, the results in a series of three-dimensional multi-objective optimization for a single and multiple pin fins between two parallel walls are presented in Sec. 3.3.

### 3.1. 2D optimization for drag reduction

Shape optimization of a 2D cylinder in a uniform inflow for reducing drag is considered as a benchmark problem for verifying the present algorithm. This problem has been tackled in previous studies with different approaches, including deep neural networks [19] and adjoint methods [5], [6], etc. We follow the computational setup of the direct numerical simulation (DNS) conducted by Park et al. [35]. Specifically, a single circular cylinder is placed in the center of the computational domain, which is 100 times the cylinder diameter ($d$) in both the streamwise and lateral directions, as shown in Fig. 4(a). No-slip and iso-thermal conditions are imposed on the cylinder surface through the VPM terms in Eqs. (6) and (8). Dirichlet and Neumann boundary conditions for the velocity and the temperature are used at the inlet and outlet boundaries, respectively. The top and bottom surfaces are treated as free-slip and impermeable boundaries. The Reynolds number is defined based on the inlet velocity and the cylinder diameter. In this study, two Reynolds numbers, $Re$ = 10 and 40, are considered for the comparison with the results reported in the previous studies [5], [6], [19].

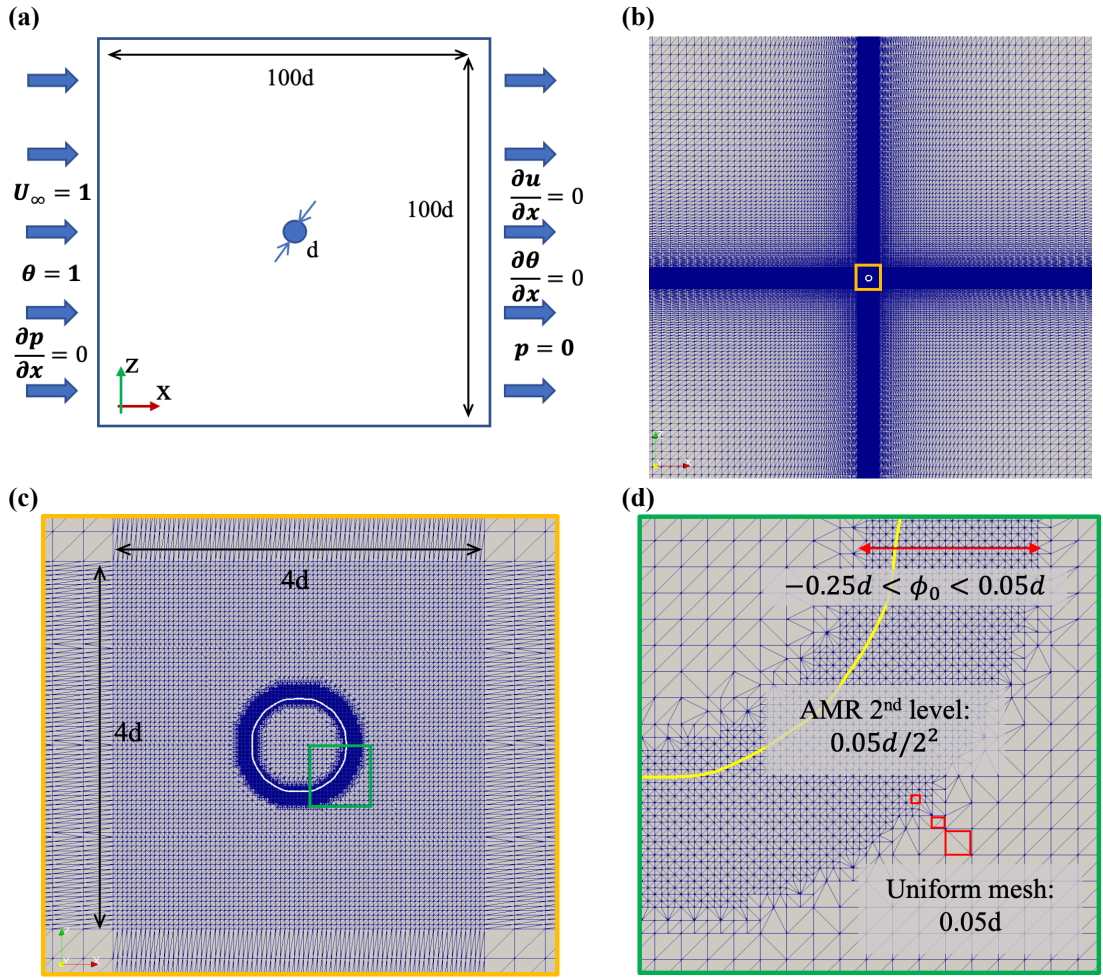

**Figure 4.** (a) Computational domain and boundary condition, (b) grid distribution in the entire domain ($100d \times 100d$), (c) enlarged view near the cylinder ($4d \times 4d$), and (d) further enlarged view of the refined region ($-0.25d < \phi_0 < 0.05d$) where AMR is applied. Here, $d$ is the diameter of the cylinder.

In the near-cylinder region with the dimensions of $4d \times 4d$, moderately fine Cartesian grids are uniformly distributed as shown in Fig. 4(c). Outside this region, the grid size gradually coarsens with a growth ratio of 1.2 as depicted in Fig. 4(b). The second-level AMR is applied based on the level-set function in the near-interface region of $-0.25d < \phi_0 < 0.05d$, as shown in Fig. 4(d). The refinement region adaptively follows the interface during the optimization by Step 7 in Fig. 3. The Cartesian grids inside the near-cylinder region with the dimension of $4d \times 4d$ are systematically changed as 0.08d, 0.05d, and 0.03d, which are hereafter referred to as coarse, medium and fine meshes, as shown in Table 1 and Fig. 5, while the level of AMR is kept constant. Since AMR automatically slits the uniform mesh, the grid resolution near the fluid-solid interface is also refined accordingly. Table 1 shows that the drag coefficients of the initial shape at $Re$ = 10 and 40 obtained by the medium mesh has relative errors less than 0.5% and 0.8%, respectively, in comparison to the DNS results reported by Park et al [35]. Therefore, it can be concluded that the medium grid resolution provides sufficient accuracy for the forward simulation.

In the current 2D problem, the total drag force is defined as $Fo_i = \int_\Omega \phi \eta_u u_i dV$. Therefore, the cost function (9) reduces to $J = C_{F_i} F_i$, where $C_{F_i} = (1,0,0)$ is set to reduce the drag in the streamwise direction. The drag coefficient is defined as $C_d = F_x/(0.5\rho U_\infty^2 d)$. In the adjoint analysis, the boundary conditions for the adjoint variables are derived so as to cancel all the boundary terms from Eqs. (18a) to (18e). Specifically, at the inlet, the adjoint velocity is fixed to zero, while a zero-gradient condition is imposed to the adjoint pressure. At outlet, the adjoint velocity is implicitly determined by $Re u_j^* u_1 + \partial u_j^*/\partial x_1 = 0$, while the adjoint pressure is set to be $p^* = u_i^* u_i$. Similar to the forward simulation, the top and bottom boundaries are treated as free-slip and impermeable boundaries.

Figure 5 shows the evolution of the cost functional as a function of the iteration number. The present cost functionals monotonically decrease and converge to lower values than those reported by Chen et al. [19] based on deep neural networks for both $Re$ = 10 and 40. Please note that the volume of the solid object is kept constant by scaling the level-set function after reinitialization so as to avoid the complete removal of the solid. The evolution curves obtained by different grid resolutions are also compared, and the results with the medium grid resolution show good convergence as shown in Table 1.

**Table 1.** Comparison of the drag coefficients for the initial and optimal shapes obtained in the present study with the reference data of Chen et al. [19], Kim and Kim [5], and Park et al. [35] at $Re$ = 10 and 40.

|  | *Present results* |  |  | *Reference data* |  |  |
| --- | --- | --- | --- | --- | --- | --- |
|  | Coarse grid | Medium grid | Fine grid | Chen et al. (2021) [19] | Kim & Kim (1995) [5] | Park et al. (1993) [35] |
| **Initial** | 2.774 | 2.768 | 2.741 | 2.740 | 2.720 | 2.780 |

| | | | | | | | |
|---|---|---|---|---|---|---|---|
| $Re = 10$ | Optimal | 2.661 | 2.463 | 2.444 | 2.501 | 2.373 | N/A |
| | Reduction rate | 4.07% | 11.02% | 10.84% | 8.72% | 12.11% | N/A |
| $Re = 40$ | Initial | 1.521 | 1.499 | 1.501 | 1.396 | 1.462 | 1.510 |
| | Optimal | 1.342 | 1.143 | 1.149 | 1.228 | 1.111 | N/A |
| | Reduction rate | 11.77% | 23.75% | 23.45% | 12.03% | 24.01% | N/A |

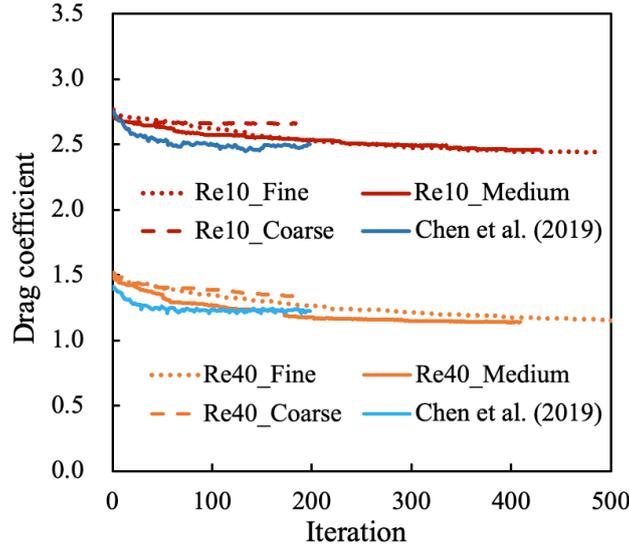

**Figure 5.** The evolution of drag coefficient as a function of an iteration number for drag reduction problem of a 2D cylinder with different grid resolutions. The reference data reported by Chen et al. [19] at $Re$ = 10 and 40 are also plotted for comparison.

The comparison of the optimal shapes between the present results and those reported in the literature [5], [19] at the Reynolds numbers of 10 and 40 is shown in Fig. 6(a) and 6(b), respectively. In the present optimization for $Re$ = 10, the convergence criterion (22) is met after 430 forward-adjoint iterations. The present optimization leads to a streamlined shape with a sharp trailing edge, which achieves a lower value of the drag coefficient than the similar optimal shape reported by Chen et al. [19]. Optimization results after 215 and 700 iterations are also shown together with that obtained after 430 iterations in Fig. 6(a). The optimal shape after 215 iterations shown in Fig. 6(a) and the corresponding drag coefficient shown in Fig. 5 both agree well with the reference results reported by Chen et al [19]. The present results indicate that continuing the forward-adjoint simulation after 215 iterations leads to further drag reduction in the present study. The optimal shape obtained after 700 forward-adjoint iterations in the present study is in better agreement with the optimal shape obtained by Kim and Kim [5]. It should be noted that, however, the decrease of the drag coefficient after 430 forward-adjoint iterations is minor. As the shape elongates in the streamwise direction, the pressure drag decreases due to a smaller projected area, while the viscous drag increases due to a larger wetted area. Hence, the improvement of the overall drag is saturated even though the shape is updated after 430 iterations.

(a)

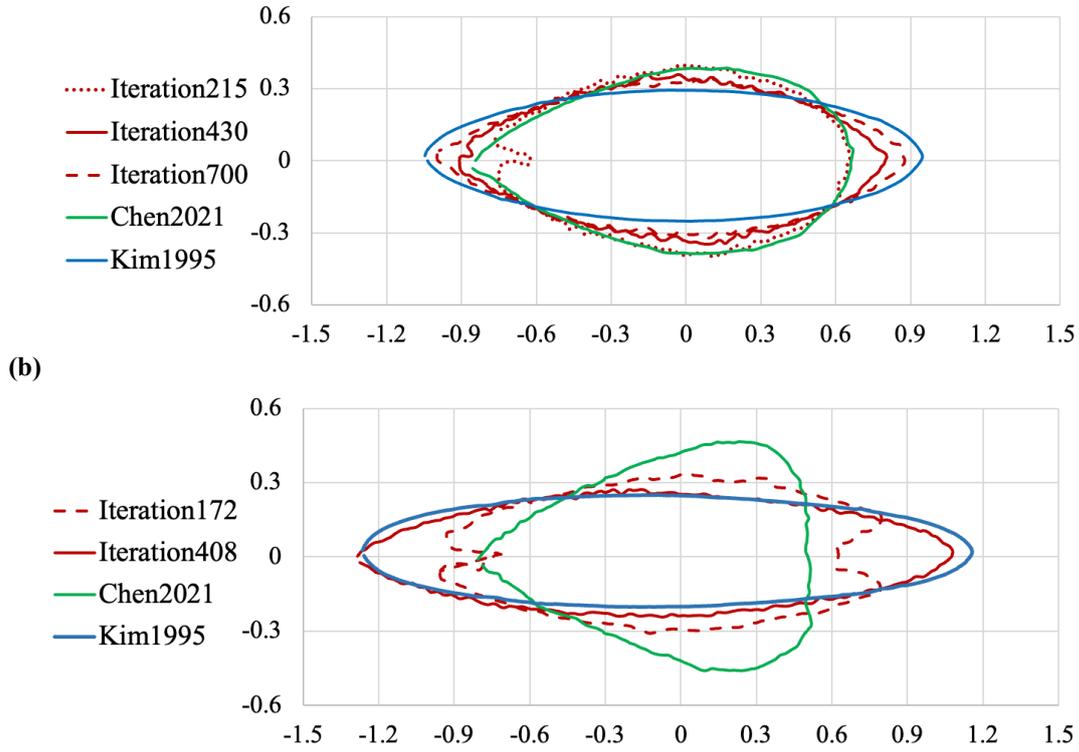

**Figure 6.** Comparison of the optimal shapes obtained in the present study at different iteration numbers with those reported by the previous studies of Chen et al. [19], and Kim and Kim [5] for the two different Reynolds numbers of (a) $Re = 10$ and (b) 40.

For $Re = 40$, the cost functional converges after 408 iterations, and the optimal shape again agrees well with that reported in Kim and Kim [5], while there exist clear differences from the bullet shape obtained by Chen et al. [19], as shown in Fig. 6(b). The comparison of the drag coefficients obtained in the present study and also the reference studies is summarized in Table 1. The disagreement may stem from the fact that the present study uses the level-set method to represent the shape, while Chen et al. [19] used a Bézier curve with a limited number of control points, so the degrees of freedom are much lower than the present study. In addition, the present study and Kim and Kim [5] directly solve the adjoint equations for obtaining the sensitivity, while Chen et al. [19] uses deep neural network to approximate the sensitivity.

Next, the effect of the initial shape on the optimization results is also investigated at $Re = 10$. Specifically, we consider a square cylinder oriented with an angle of 45 degrees with respect to the flow direction as an initial shape as shown with a blue dashed line in Fig. 7 (a), while the optimal shape is depicted by a solid blue line. The optimization results obtained from the initial circular cylinder are also plotted by red for comparison. The two optimization cases start from the same solid volume, and ends at the same iteration number of 430. A similar shape, but relatively thicker in the lateral direction is obtained from the square cylinder than that from the circular cylinder. It is unclear whether these two cases eventually converge to a similar shape after a sufficient number of iterations. According to the evolution of the cost functionals shown in Fig. 7 (b), however, it is unlikely because the reduction of the drag coefficient for the square cylinder is 2.55 after 430 iterations and it starts converging, while that for the circular cylinder is

2.46 and still decreasing. The present results suggest that there exists the effect of an initial shape on the optimization results, and this is reasonable considering that the present scheme is a gradient-based method and therefore there is a possibility that a solution is stuck in a local minimum.

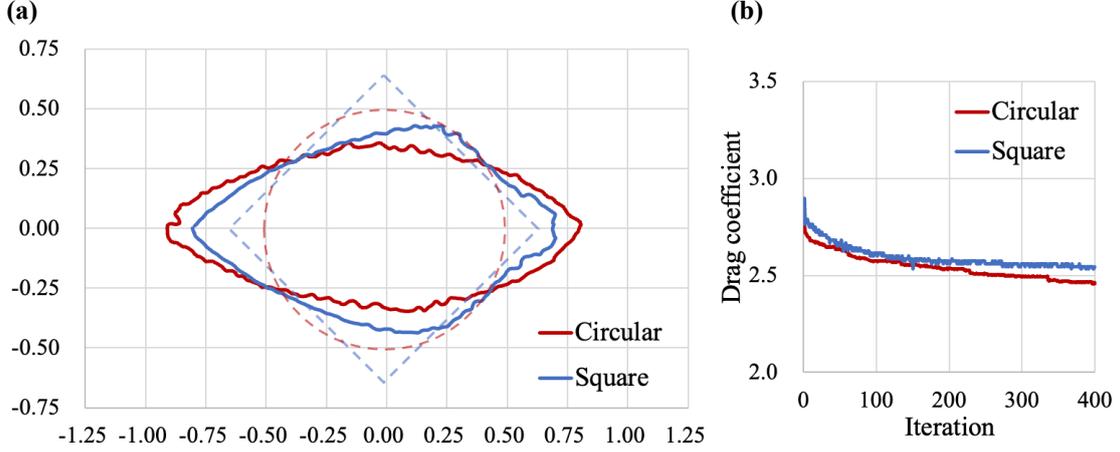

**Figure 7.** Effects of an initial shape on the optimization results at *Re* = 10. (a) Optimal shapes (solid lines) evolve from initial (light dashed lines) circular and square shapes. (b) Evolutions of the drag coefficient as a function of an iteration number.

### 3.2. 2D optimization for heat transfer enhancement

Here, we extend the 2D drag reduction problem considered in the previous subsection by taking into account heat transfer enhancement as well. Specifically, we set the cost functional (9) as $J = C_Q \tilde{Q} + C_{F_x} \tilde{F}_x$, where $\tilde{Q}$ and $\tilde{F}_x$ are the heat transfer and the drag force on the unit length of a square cylinder, normalized with their initial values. Here, we focus on the case, in which $C_{F_x} = 0.89$ and $C_Q = -1$. The same square cylinder presented in Fig. 7 is employed as the initial shape.

The computational domain, the mesh, and the single square pin fin are set as same as the drag reduction cases demonstrated in Fig. 4. Dirichlet and Neumann boundary conditions are also respectively imposed at the inlet and outlet boundaries to solve the energy equation (8). Adiabatic conditions are imposed at the top and bottom boundaries. The Reynolds number based on the cylinder diameter is maintained as 10. In order to show the optimal shapes in different working fluids, the Prandtl number is set to be 6.9 and 0.7, which corresponds to water and air, respectively, under the standard condition. As for the adjoint temperature boundary conditions, $\theta^*$ is fixed to zero at the inlet, while it is implicitly determined by $PrRe\theta^* u_1 + \partial \theta^* / \partial x_1 = 0$ at the outlet similar to those for the adjoint velocity.

The grid dependency on the Nusselt number and the drag coefficient is carried out for the initial square shape as shown in Table 2. The Nusselt number of the 2D pin fin is defined as

$$Nu = \frac{QL}{\lambda_f A(\Theta_b - \Theta_w)}, \qquad (23)$$

where $L = d$ as we set the diameter as the characteristic length, $A$ is the surface area of the unit height of the square fin, and the solid temperature $\Theta_w$ is zero due the thermal boundary condition. It can be confirmed that the medium mesh is sufficient for evaluating the drag and the overall heat transfer at both the Prandtl numbers.

Table 2. Nusselt number and drag coefficient of an initial square cylinder at $Re = 10$ for $Pr = 6.9$ and 0.7 obtained with the same medium and fine grids in Sec. 3.1.

|  | Medium | | Fine | |
|---|---|---|---|---|
| $Pr = 6.9$ | $Nu = 3.60$ | $C_d = 3.04$ | $Nu = 3.57$ | $C_d = 3.02$ |
| $Pr = 0.7$ | $Nu = 1.82$ | | $Nu = 1.81$ | |

The evolutions of the cost functional, overall drag and heat transfer as a function of the iteration number for $Pr = 6.9$ and 0.7 are shown in Fig. 8(a) and (b), respectively. Note that we did not run the optimization until the cost functional meets the convergence criterion (22) because of the increasing structure complexity for heat transfer enhancement in the present configuration. Hence, the iteration number is fixed to 25. It can be seen that the cost functional monotonically decreases, supporting the validity of the present scheme. The effects of the grid resolution are relatively minor but tend to be larger with increasing the forward-adjoint iteration. This is due to the complex structures of the fluid-solid interface in the optimal shapes as will be explained later. The optimization history shows that the heat transfer is enhanced by over 10% in both cases, while the drag force is increased by 5% for $Pr = 6.9$ and is reduced by 1.2% for $Pr = 0.7$ from that of the initial shape.

Figure 9 shows the optimal shapes obtained after 25 forward-adjoint iterations with different grid resolutions. In general, similar optimal shapes are obtained with different resolutions. Specifically, the top and bottom vertexes are smoothened to reduce drag, while the upstream vertex extends towards the flow direction to further enhance heat transfer. They are evident when compared with the optimal shape considering drag reduction only at the same iteration number of 25 as shown in the red solid line in Fig. 9(a). More careful observation reveals that distinctive roughness is generated on the inclined upstream surfaces for higher Prandtl number of 6.9, and it appears only for the higher grid resolution. As will be discussed below, the small roughness is effective for heat transfer enhancement at the higher Prandtl number, and the finer mesh is required to resolve such a roughness.

(a)                                      (b)

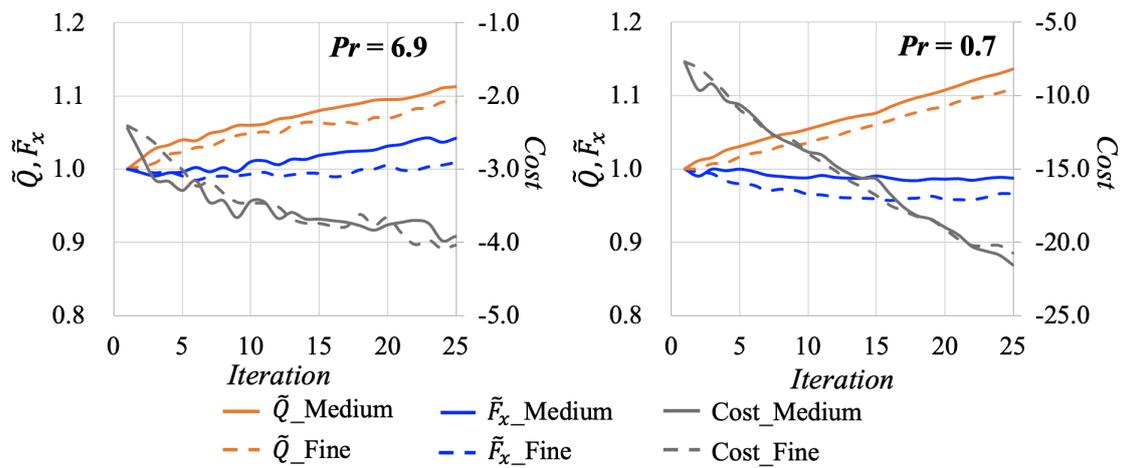

**Figure 8.** Evolutions of the normalized heat transfer ($\tilde{Q}$), the normalized drag force ($\tilde{F}_x$), and the cost functional using different mesh for (a) $Pr = 6.9$ and (b) $Pr = 0.7$.

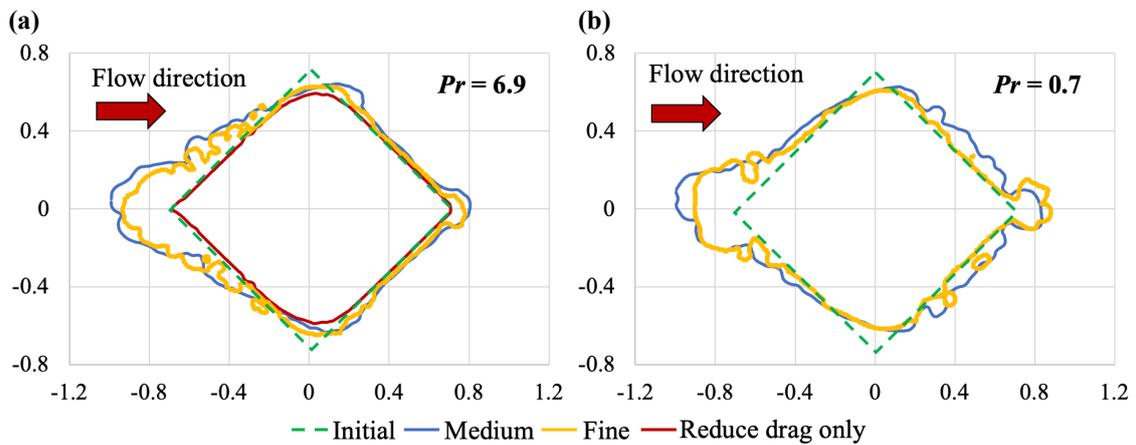

**Figure 9.** Optimal shapes for enhancing heat transfer and minimizing drag at $Re = 10$ obtained with the medium and the fine grid resolutions for (a) $Pr = 6.9$ and (b) $Pr = 0.7$. The red line in (a) shows the optimal shape obtained for drag reduction only at the same iteration number of 25.

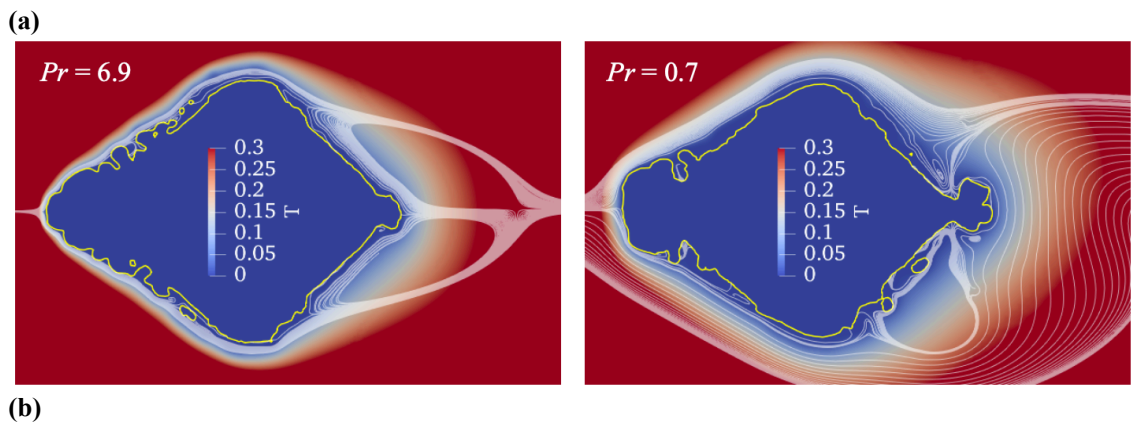

(a)

(b)

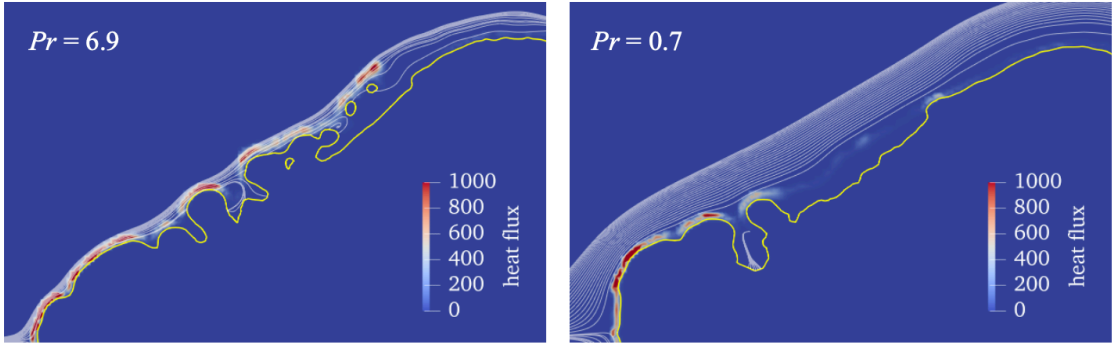

**Figure 10.** Spatial distributions of (a) temperature and (b) heat flux at the fluid-solid interface in the optimal shapes for (left) $Pr = 6.9$ and (right) 0.7. The white curves show streamlines around the optimal shapes

The thermal and velocity fields (shown by streamlines) around the optimal shapes for $Pr = 6.9$ and 0.7, and also the expanded views of the artificial heat source $\eta_\theta \phi \theta$ in energy equation (8) corresponds to a local wall heat flux are shown in Fig. 10. For $Pr = 6.9$, the local peaks of high heat fluxes are observed in the convex parts, while the heat flux for $Pr = 0.7$ has its maximum around the upstream stagnant point rather than the inclined surface. This could be attributed to a thinner thermal boundary layer formed along the inclined upstream surface at the higher Prandtl number of 6.9, and explains why the characteristic roughness emerges on the inclined surface at $Pr = 6.9$.

### 3.3. 3D multi-objective optimization of a single pin fin in channel flow

Next, we further extend the current framework to three-dimensional shape optimization problems. As a representative heat transfer surface, we consider a single square pin fin sandwiched between two parallel walls as an initial geometry as shown in Fig. 11. The origin of the coordinate system is fixed at the center of the square pin fin on the bottom surface. The channel height is taken as the reference length-scale, and the edge length of the square is identical to the channel height. The streamwise, wall-normal, and spanwise directions are denoted by $x$, $y$ and $z$, respectively. The dimensions of the computational domain in the $x$, $y$, and $z$ direction are 5.0, 1.0, and 2.5, respectively. The center of the square cylinder is located at a distance of 2.0 from the inlet boundary and 1.25 from the side boundaries Note that, similar to the previous sections, the 3D square cylinder is oriented with an angle of $45°$ with respect to the freestream direction as shown in Fig. 11.

In the present configuration, drag and heat transfer occur not only on the surface of the pin fin, but also the iso-thermal two parallel walls connected to the fin. Therefore, the wall friction ($Fb_i$) on the channel walls (i.e., the top and bottom boundary in Fig. 11) is also included into the total drag force ($F_i$) in the cost functional Eq. (9), while the wall heat flux ($Qb$) on the two walls is also taken into account in the total heat transfer ($Q$). Accordingly, the total drag and the total heat transfer are respectively defined as

$$F_i = Fo_i + Fb_i = \int_\Omega \phi \eta_u (u_i - u_i^s) dV + \int_{d\Omega_{wall}} \left\{ \frac{-1}{Re} \left( \frac{\partial u_i}{\partial x_j} + \frac{\partial u_j}{\partial x_i} \right) n_j \right\} dS, \qquad (24)$$

$$Q = Qo + Qb = \int_{\Omega} \phi \eta_\theta (\theta - \theta^s) dV + \int_{d\Omega_{wall}} \frac{-1}{PrRe} \frac{\partial \theta}{\partial x_j} n_j dS. \qquad (25)$$

where $n_j$ is the outward unit normal vector on the boundary wall ($d\Omega_{wall}$). Here, the drag on the solid object ($Fo_i$) and the heat transfer on the pin-fin interface ($Qo$) are the same as those in Eq. (10) and (11).

The boundary conditions for physical quantities in the forward simulation are shown in Table 3. A periodic boundary condition is imposed in the spanwise direction. No-slip and iso-thermal conditions, i.e., $u_i = \theta = 0$, are imposed on the top and bottom walls. Uniform velocity and temperature of $U_\infty = 1$ and $\theta_\infty = 1$ are adopted at the inlet, while a Neumann boundary condition is applied to the outlet boundary. The Reynolds number based on the channel height and the inlet velocity is set to be $Re = 10$, while the Prandtl number is set to be 6.9. Due to the low-Reynolds number, the velocity and thermal fields are assumed to be steady.

Table 3 also summarizes all the boundary conditions for the adjoint variables, which are derived so as to cancel all the boundary terms from (18a) to (18e). The adjoint boundary conditions are generally similar to those of the forward equations. In particular, at outlet, the adjoint variables are implicitly given and iteratively updated based on the forward fields. The inlet and outlet boundary conditions are exactly the same as the 2D problems in the previous sections. However, adding $Fb_i$ and $Qb$ into the cost functional for the 3D problems bring two extra boundary terms in Eq. (14), which determine the adjoint velocity and thermal boundary conditions for the top and bottom boundaries as shown in Table 3.

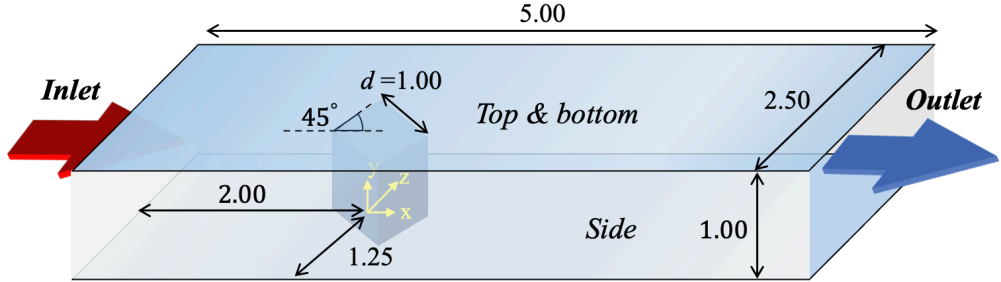

**Figure 11.** Computational domain and coordinate system for three-dimensional pin-fin optimization.

**Table 3.** Boundary conditions for forward and adjoint simulations

| | *Inlet* | *Outlet* | *Top & bottom* | *Side* |
|---|---|---|---|---|
| Velocity ($u_j$) | $u_j = (1,0,0)$ | $\frac{\partial u_i}{\partial x_j} n_j = (0,0,0)$ | $u_j = (0,0,0)$ | |
| Temperature ($\theta$) | $\theta = 1$ | $\frac{\partial \theta}{\partial x_j} n_j = 0$ | $\theta = 0$ | |
| Pressure ($p$) | $\frac{\partial p}{\partial x_j} n_j = 0$ | $p = 0$ | $\frac{\partial p}{\partial x_j} n_j = 0$ | Periodic |
| Adjoint velocity ($u_j^*$) | $u_j^* = (0,0,0)$ | $u_j^* u_1 + \frac{1}{Re}\frac{\partial u_j^*}{\partial x_1} = 0$ | $u_j^* = C_{Fb_j} V_\Omega$ | |
| Adjoint temperature ($\theta^*$) | $\theta^* = 0$ | $\theta^* u_1 + \frac{1}{Pe}\frac{\partial \theta^*}{\partial x_1} = 0$ | $\theta^* = C_{Qb} V_\Omega$ | |
| Adjoint pressure ($p^*$) | $\frac{\partial p^*}{\partial x_j} n_j = 0$ | $p^* = u_i^* u_i + \theta^* \theta$ | $\frac{\partial p^*}{\partial x_j} n_j = 0$ | |

In the present multi-objective TO problem, we set a cost functional as $J = C_{F_i}\tilde{F}_i + C_Q\tilde{Q}$, where

$$\tilde{F}_i = F_i/F_{i,0} \text{ and } \tilde{Q} = Q/Q_0 \tag{26}$$

are normalized by their initial magnitude. Here, the subscript of 0 indicates an initial value. $C_{F_i} = (C_{F_x}, 0, 0)$ is considered for reducing the drag in the streamwise ($x$) direction only. In this study, we investigate seven different combinations of the weight coefficients $C_{F_x}$ and $C_Q$, and each case is named based on the values of the weight coefficients as summarized in Table 4. Since we aim to enhance heat transfer, $C_Q$ is set to be negative, whereas $C_{F_x}$ is positive to reduce the total drag. In all the cases, the volume of the fin is kept constant to avoid a trivial solution, such as complete removal of the solid object for reducing drag in Case F10Q0, and flow blockage for enhancing heat transfer in Case F0Q10. On the other hand, a relatively large weight for heat transfer enhancement, e.g., Cases F0Q10 or F1Q9, results in quite complicated structure, so that it is difficult to continue until the convergence criterion (22) is satisfied. Therefore, we fix the iteration number to 25 and compare the resultant heat transfer and pressure drop characteristics in all the 3D cases.

Table 4. Cases considered for single three-dimensional pin-fin optimization.

| Case | F10Q0 | F9Q1 | F7Q3 | F5Q5 | F3Q7 | F1Q9 | F0Q10 |
|---|---|---|---|---|---|---|---|
| $C_{F_x}$ | 1.0 | 0.9 | 0.7 | 0.5 | 0.3 | 0.1 | 0.0 |
| $C_Q$ | 0.0 | -0.1 | -0.3 | -0.5 | -0.7 | -0.9 | -1.0 |

First, we confirm that the cost functionals decrease monotonically with increasing the iteration number in all the cases as shown in Fig. 12(a). The variations of the total drag and heat transfer as a function of the iteration number in each case are also shown in Fig. 12(b) and (c), respectively. When the weight of heat transfer is relatively small, the total drag is kept smaller than that of the initial shape, while the total heat transfer is enhanced (see, Cases F3Q7, F5Q5 and F7Q3). Specifically, Case F3Q7 enhances the total heat transfer by 4.0% and simultaneously reduces the pressure drop by 12.6% from that of the initial shape. In Cases F1Q9 and F0Q10, the weight for heat transfer is relatively large, so that the optimal shapes enhance heat transfer significantly at the expense of an increased pressure loss.

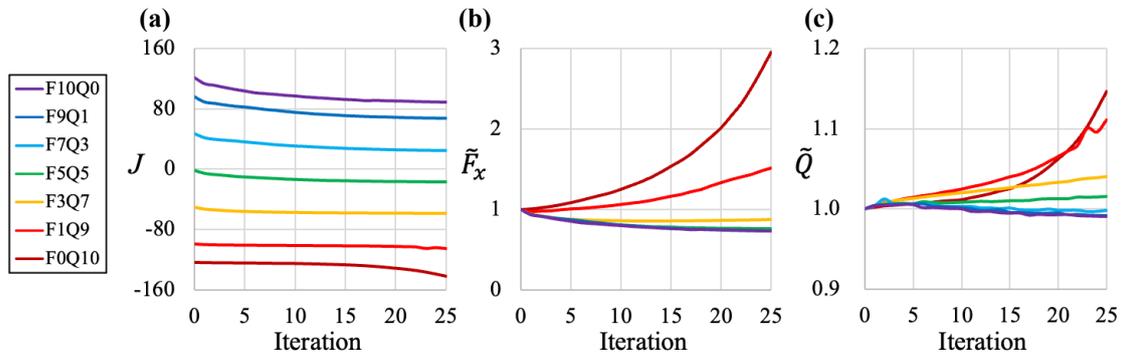

**Figure 12.** (a) Optimization history of cost functional, (b) normalized total drag force in streamwise direction, and (c) normalized total heat transfer.

Figure 13 shows the trajectory of each optimization case in the two-dimensional plot of the total drag and heat transfer for 25 iterations. All the cases start from the same initial shape corresponds to $(\tilde{F}_x, \tilde{Q}) =$ (1.0, 1.0). It can be seen that the updated shapes with the different weight coefficients for the drag and heat transfer move toward the left and/or top direction to form the Pareto front.

Figure 14 shows the optimal shapes obtained after 25 iterations in all the cases with the local sensitivity indicated by colors on the pin-fin surfaces. In Cases F10Q0 and F9Q1, the projected area of the optimal shape on the streamwise cross plane decreases to reduce the drag. Their shapes look similar to the streamwise-elongated shape obtained in the 2D optimization case for drag reduction in Sec. 3.1. However, strong non-uniformity in the wall-normal direction can be confirmed in the present three-dimensional cases. In Case F7Q3, a protrusion appears on the upstream surface of the fin around the channel center, where the prominent positive sensitivity can be confirmed. With further increasing the weight for heat transfer, the number of protrusions is increased, and each protrusion extends toward the upstream direction. These results suggest that such convex structures are effective in enhancing heat transfer. Finally, large protrusions are extended to the spanwise direction for Cases F1Q9 and F0Q10, where the weight coefficient for heat transfer is sufficiently large to overcome the penalty of the drag.

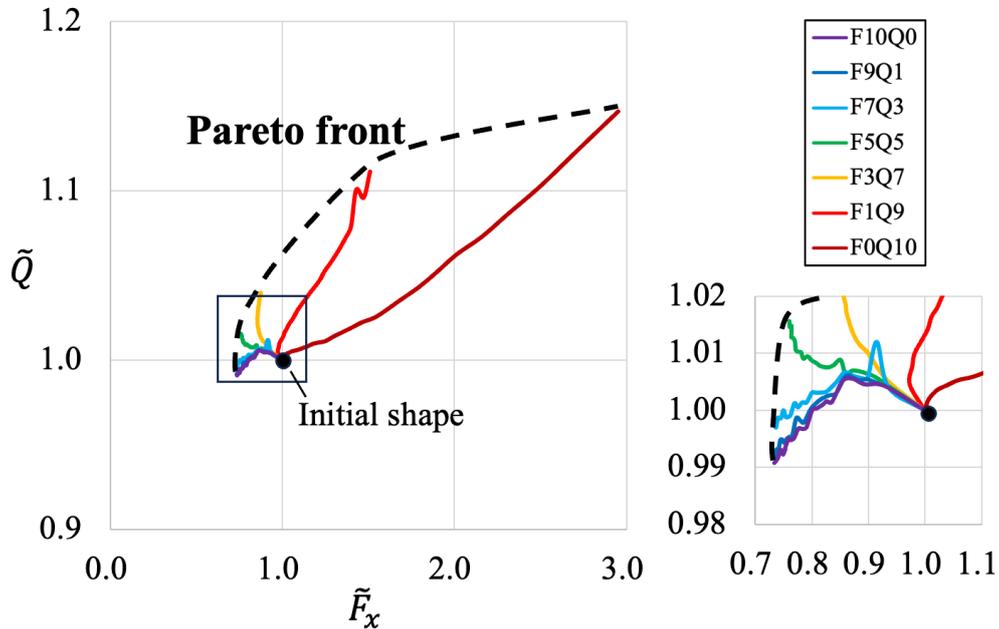

**Figure 13.** Evolutions of optimization results in 2D map of the total streamwise drag (horizontal axis) and heat transfer (vertical axis) during the first 25 iterations with different weight coefficients for heat transfer and pressure drag.

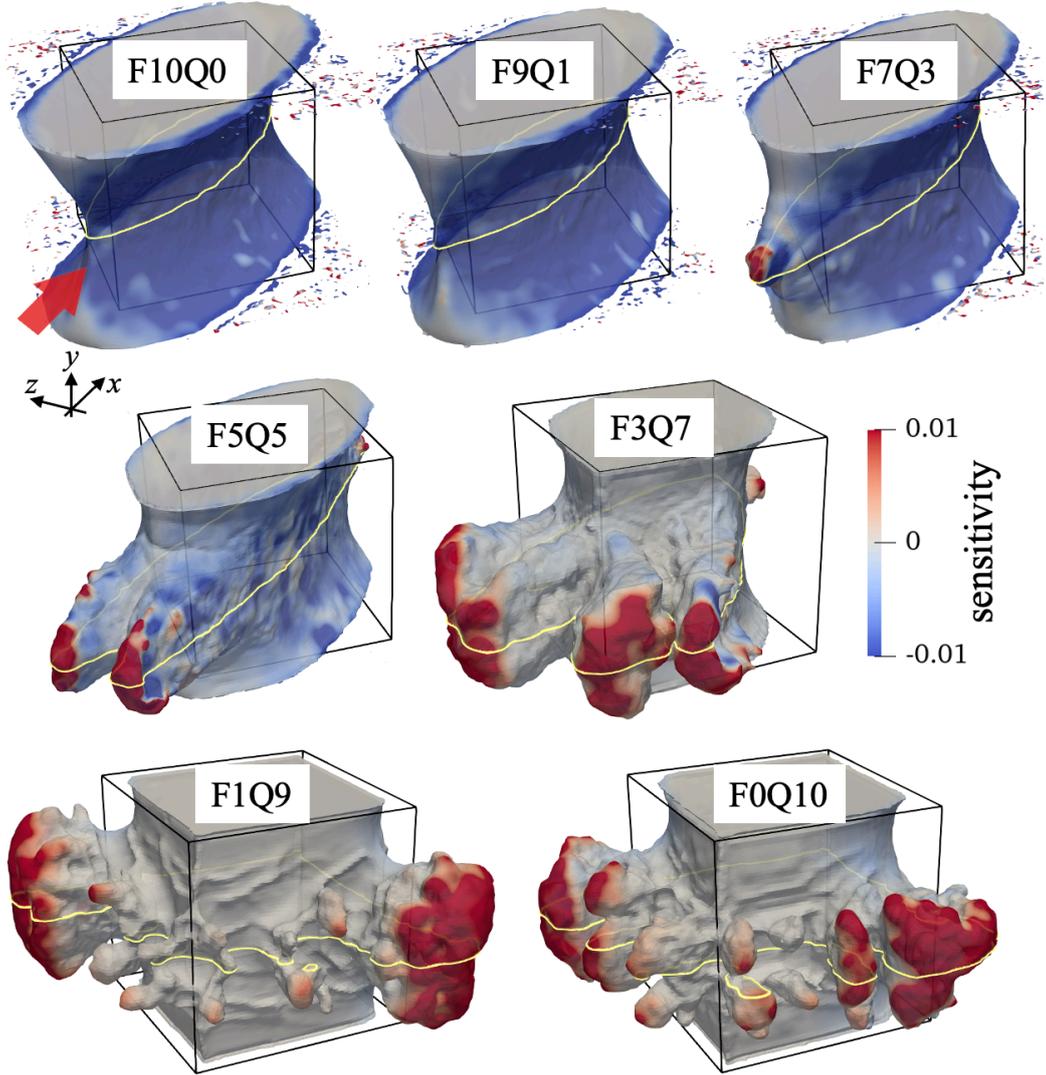

**Figure 14.** Optimal shapes obtained with different weights for drag and heat transfer after 25 iterations. The local sensitivity on each surface is shown with colors from blue to red. Positive sensitivity indicates the outward surface displacement, while negative sensitivity is the inward displacement. The yellow line depicts the cross section of the optimal shape at the channel center. The flow direction is from left-bottom to right-top as shown by the red arrow in Case F10Q0. The initial square cylinder is depicted by the solid lines in each figure.

### 3.4. 3D optimization on pin fin array in channel flow

Finally, we showcase the application of the present framework to topology optimization problem of a 3D pin-fin array. As for the initial shape, we consider six circular pin fins placed in a staggered arrangement between two parallel plates as illustrated in Fig. 15. The dimensions of the computational domain are 6.0, 1.0, 2.0 in the *x, y, z* directions, respectively. The diameter of each pin fin is 0.5, and the spacing is 1.0. Similar to the previous single pin fin cases, the channel height is chosen as the reference length, while the inlet velocity is the reference velocity scale. The Reynolds number and Prandtl number are respectively set to be 10 and 6.9 as in the single pin fin case discussed in Sec. 3.3. The boundary conditions are also maintained consistent with the previous cases. The cost functional is set to be $J = C_{F_x}\tilde{F}_x + C_Q \tilde{Q}$ for enhancing total heat transfer while suppressing the total streamwise drag. Here, the weight coefficients are set as $C_{F_x} = 0.1$, $C_Q = -0.9$.

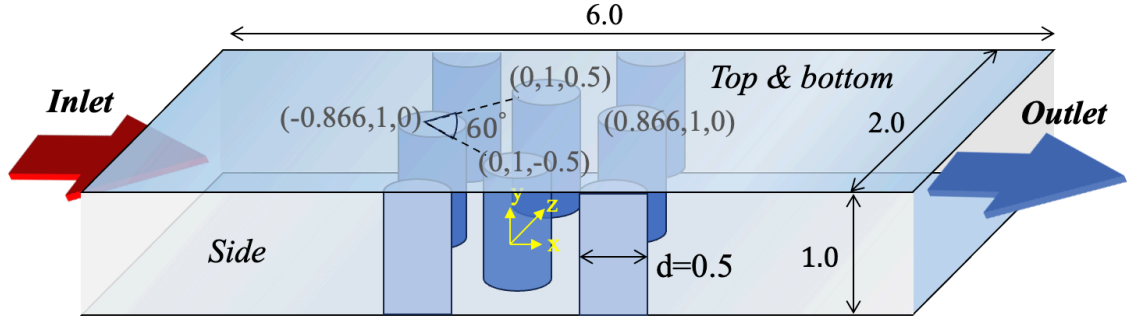

**Figure 15.** Computational domain and coordinate system for the topology optimization of a pin-fin array sandwiched between two parallel walls. The coordinates of the centers of the cylinders on the top surface are given. The origin of the coordinate locates at the center of the bottom plate.

The evolutions of the cost functional, the total drag and heat transfer are plotted as a function of the iteration number in Fig. 16 (a). Figure 16(b-d) shows the evolution of the pin-fin array and the corresponding temperature field on the middle plane of the channel at iteration numbers of 20, 80, and 150. Note that we duplicate the same structure in the spanwise direction for better visibility considering the periodic boundary condition applied in this direction. The optimal shape has a 19% increase in total heat transfer, and also a 8.7% decrease in total drag from that of the initial shape after 150 iterations. The monotonic reduction of the cost functional with the number of iterations again verifies the present optimization code. Similar to the cases of a single fin, the secondary fins (or protrusions) grow on the upstream side of the pin fins around the middle of the channel after the first 20 iterations. The emergence of the secondary fins significantly enhances the heat transfer as can be seen in the change of $\tilde{Q}$ in Fig. 16(a). After 30 iterations, however, the pin fins in each row are updated in different manners. Specifically, the pin fins in the second row start shrinking and eventually disappear after 30 iterations (see, Fig. 16 (c) and (d)), while the fins in the first rows continue extending toward the streamwise direction with a certain angle to the streamwise direction (see, structures A and B in Fig. 16 (d), and their expanded views in Fig. 16 (e)). Such a unique structure contributes to not only heat transfer by itself, but also guiding a high-temperature fluid to the third row of the pin fins for further heat transfer enhancement as can be seen in Fig. 16(c) and (d). Accordingly, secondary fins also grow on the pin fins in the third row so as to effectively extract heat from high-temperature fluid distributed from the fins in the first row (see, structure C in Fig. 16 (d), and its expanded view in Fig. 16 (e)). It is considered that the first and third rows of the pin fins have sufficient heat transfer performances, and the second row contributes to pressure drop more than heat transfer. That is the reason why the second row of the pin fins disappeared during the optimization process. In summary, the present scheme can handle such complex structures with topology change in a straightforward manner, and successfully optimize complex three-dimensional structures for simultaneous achievement of heat transfer enhancement and drag reduction.

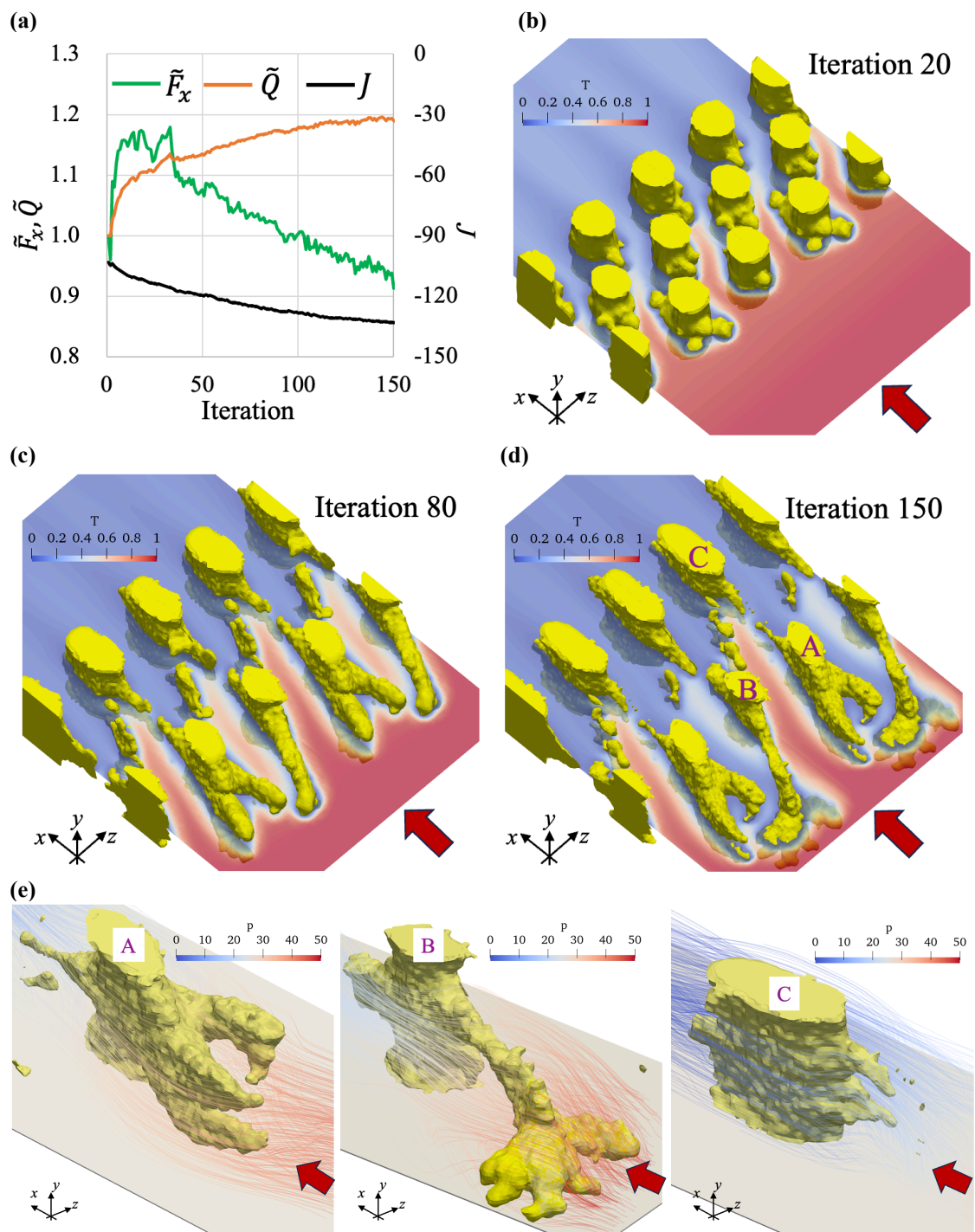

**Figure 16.** (a) Cost functional, total drag force, and total heat transfer versus the iteration number. 3D visualization of the optimal pin-fin array and the temperature on the middle plane after (b) 20, (c) 80, and (d) 150 iterations. (e) Expanded view of the structures of the fins denoted by A, B and C in (d). Lines represent streamlines and their color indicates local pressure. In all the figures, the flow direction is shown by the red arrow.

## 4. Conclusion

We proposed an efficient and robust topology optimization algorithm for heat transfer enhancement and drag reduction by combining the adjoint analysis with the level-set method to embed an arbitrary three-dimensional fluid-solid interface in a Cartesian grid system. The newly proposed algorithm for the reinitialization procedure of the level-set function allows to suppress the numerical diffusion of the interface and ensure the interface is maintained within a local single grid spacing. Based on the level-set function, adaptive mesh refinement effectively allocates fine computational grids only near the fluid-solid interface to achieve high accuracy at a reasonable computational cost. The overall numerical codes including forward and adjoint analyses as well as the update and reinitialization of the level-set function were successfully implemented in an open-source CFD software OpenFOAM.

We first validated the present optimization algorithm in the shape optimization problem of a 2D cylinder placed in a uniform flow. The drag coefficients of the initial circular cylinder at Re = 10 and 40 obtained in the present solver showed good agreement with those obtained by DNS with a body-fitted mesh reported in the previous study [35]. In addition, the optimal shapes and the corresponding drag coefficients are also compared with the previous optimization results. It was confirmed that the present level-set based optimization code result in similar or even better results that those reported by Chen et al. [19]. It should be noted that these previous studies employed a body-fitted mesh, so that they cannot easily handle topology changes, while the present approach allows to embed arbitrary three-dimensional geometries in a Cartesian mesh, and therefore deals with topology change in a straightforward manner. Next, we applied the present algorithm to a multi-objective optimization problem where both heat transfer enhancement and drag reduction are taken into account for two different working fluids, i.e., air and water with Pr = 0.7 and 6.9, respectively, at Re = 10. As a result, heat transfer is enhanced by 10%, while the pressure drag is almost unchanged or even reduced by around 5%.

Next, the present algorithm was extended to three-dimensional optimization problems for single and multiple square fins between two parallel plates for the overall heat transfer enhancement and the overall streamwise drag. For the former case of a single fin, it can be confirmed that, with increasing the relative weight for heat transfer enhancement, the optimal fin changes from a streamlined elliptical structure to that with more complex secondary fins on the upstream surface of the primal fin. When the ratio of the weight coefficients of the normalized total drag and total heat transfer is 3:7, the optimal shape achieves 4.0% enhancement in heat transfer, while 12.6% reduction in drag. For the latter case of multiple fins, it was found that the optimal fin shapes are different in each row. Specifically, the pins in the first row extend secondary fins towards the upstream with a certain angle to the flow, and this contributes not only to heat transfer, but also to rectifying the flow, so that it impinges on the fins in the third row for further heat transfer enhancement. Meanwhile, the fins in the second row disappears to mitigate the overall pressure drop. Consequently, the prescribed cost functional decreases significantly after forward-adjoint iterations, demonstrating that the effectiveness of the proposed algorithm.

The present study assumes that a flow is steady and laminar, and also that the solid region is iso-thermal. In most real scenarios, however, flows are unsteady and turbulent. In addition, the thermal resistance inside the solid needs to be taken into account [4], [36], especially, for the complex optimal structures obtained in the present study. There exist previous studies proposing topology optimization in turbulent regimes [12], [36], [37], and combining the present algorithm with the new reinitialization algorithm for a level-set function would be an interesting extension. Especially, the present algorithm has an advantage in resolving thin velocity and thermal boundary layers formed near the fluid-solid interface in a turbulent regime. The present scheme can also be applied to a conjugate heat transfer problem in a straightforward manner, and it should be considered in future work.

**Acknowledgement**

This study is partially supported by JSPS KAKENHI Grant Numbers JP23H01339 and JP 21H05007, and also Research and Development Program for Promoting Innovative Clean Energy Technologies Through International Collaboration, the New Energy and Industrial Technology Development Organization (NEDO). Di Chen acknowledges the financial support of Outstanding Doctoral Graduate Development Fellowship in Shanghai Jiao Tong University.